\documentclass[conference]{IEEEtran}
\IEEEoverridecommandlockouts
\usepackage{booktabs}   
\usepackage{subcaption} 

\usepackage{amsmath,amssymb,amsfonts}
\usepackage{amsthm}
\usepackage{graphicx}
\usepackage{subcaption}
\usepackage{lscape}
\usepackage{textcomp}
\usepackage{xcolor}
\usepackage{braket}
\usepackage{float}
\usepackage{lipsum}
\usepackage[lined, boxed,ruled,vlined]{algorithm2e}

\newcommand{\etal}{\emph{et al.\ }}

\begin{document}

\title{SlackQ : Approaching the Qubit Mapping Problem with A Slack-aware Swap Insertion Scheme
}

\author{
    \IEEEauthorblockN{Chi Zhang\IEEEauthorrefmark{1}\IEEEauthorrefmark{3}  Yanhao Chen\IEEEauthorrefmark{2}  Yuwei Jin\IEEEauthorrefmark{2} \\ Wonsun Ahn\IEEEauthorrefmark{1}  Youtao Zhang\IEEEauthorrefmark{1}  Eddy Z. Zhang\IEEEauthorrefmark{2}\IEEEauthorrefmark{4}}
    \vspace{1.5ex}
    \IEEEauthorblockA{\IEEEauthorrefmark{1}University of Pittsburgh
    \\}
    \IEEEauthorblockA{\IEEEauthorrefmark{2}Rutgers University
    \\}
    \vspace{1.5ex}
    \IEEEauthorblockA{\IEEEauthorrefmark{3}chz54@pitt.edu
    \\}
    \IEEEauthorblockA{\IEEEauthorrefmark{4}eddy.zhengzhang@gmail.com
    \\}
}

\maketitle

\begin{abstract}
The rapid progress of physical implementation of quantum computers paved the way for the design of tools to help users write quantum programs for any given quantum device. The  physical constraints inherent in current NISQ architectures prevent most quantum algorithms from being directly executed on quantum devices. To enable two-qubit gates in the algorithm, existing works focus on inserting SWAP gates to dynamically remap logical qubits to physical qubits. However, their schemes lack consideration of the execution time of generated quantum circuits. In this work, we propose a slack-aware SWAP insertion scheme for the qubit mapping problem in the NISQ era. Our experiments show  performance improvement by up to 2.36X at maximum, by 1.62X on average, over 106 representative benchmarks from RevLib~\cite{Wille+:ISMVL08}, IBM Qiskit \cite{IBMQiskit}, and ScaffCC \cite{JavadiAbhari+:CCF14}.
\end{abstract}

\section{Introduction}
\label{sec:intro}

\begin{figure*}[t]
\centering
\includegraphics[width=0.9\textwidth]{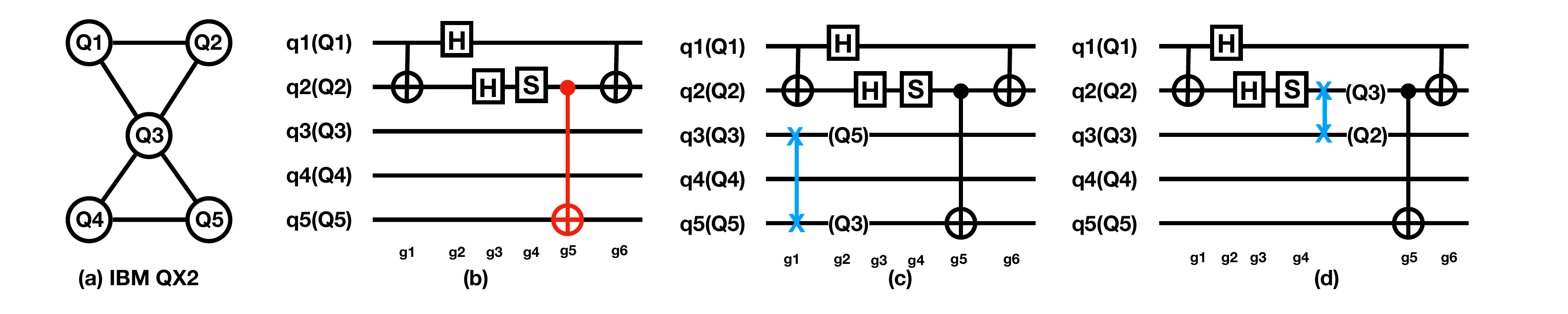}
\caption{(a) Physical qubit connectivity; (b) the original logical circuit (logical qubits q1, q2, q3, q4, q5 are mapped to physical qubits Q1, Q2, Q3, Q4, Q5);  The gate marked in red is the CNOT gate that cannot be executed due to a connectivity constraint (as Q2 and Q5 are not physically connected). (c) uses 1 swap but the execution time of the circuit is not increased; (d) uses 1 swap but the execution time of the circuit is increased.  The operations marked in blue are swap operations. We assume a swap operation can be decomposed into three CNOT gates and each gate takes 1 cycle in this example.  }
\label{fig:motivation}
\end{figure*}

Quantum computing has been considered as a potentially disruptive computation model. In 2019, Google \cite{Arute+:Nature19} demonstrated ``Quantum Supremacy" with its
54-qubit quantum processor that is able to perform a computational task in 200
seconds which would have  taken the state-of-art classical supercomputer 10,000 years. In general, quantum computing has significant advantage over classical computing for applications including large number factoring
\cite{Shor:FOCS94}, database search \cite{grover+:stoc96},
and quantum simulation \cite{peruzzo+:nature14}. 

Labs in academia and industry are now able to build quantum computers with up to 49-72 qubits. IBM~\cite{Knight:technologyreview17} released its
53-qubit quantum computer in October 2019 and has made it available for
commercial use. Google~\cite{Kelly:googleai2018} released the 72-qubit
\emph{Bristlecone} quantum computer in March 2018. Intel \cite{Hsu:CES18} and Rigetti \cite{Rigetti}
respectively have released
quantum computing devices with dozens of qubits. 
Further, a few small-scale quantum computers with less than 20 qubits are made freely available to the public \cite{IBMQ},  for example, the series of quantum computers provided by IBM Q experience \cite{IBMQ}.

The physical constraints inherent in quantum architectures prevent quantum algorithms from being directly executed on the device. One of the major constraints that must be accounted for before quantum algorithms can be  executed is the qubit connectivity constraint. In the superconducting-based quantum computers (the implementation adopted by major industry players such as IBM and Google), qubits are not fully connected. It follows nearest neighbor (NN) interaction model, enforced by the connectivity of the physical qubits array. If an algorithm requires communication between qubits that are not physically connected, the algorithm cannot be directly executed on the device.

 To solve the qubit connectivity problem, any two logical qubits that need to communicate according to the algorithm must be mapped to physical qubits on the device that are neighboring (connected). This is done through one or a sequence of SWAP operation(s). A SWAP operation exchanges the states of two neighboring qubits, in effect ``moving'' the two qubits. An example is shown in Fig. \ref{fig:motivation}. This dynamic remapping between logical and physical qubits may need to happen multiple times throughout the algorithm.     

Inserting swap operations inevitably results in increased gate count and execution time.  Previous studies \cite{Li+:ASPLOS19, Wille+:DAC19, Zulehner+:ICRC17, Zulehner+:DATE18, IBMQiskit, Siraichi+:CGO18} focus on optimizing gate count but not execution time. Execution time is an important measure of the performance of a circuit. Minimal gate count do not necessarily guarantee minimal  time. We show an example in Fig. \ref{fig:motivation} where two qubit mapping solutions yield the same gate count but only one of them is optimal in time. 

Optimizing the execution time of a circuit is important not only for optimizing the performance but also for improving the fidelity of a quantum circuit.  Quantum computers are not perfect. Qubits are fickle and error prone.  As time goes by, a qubit decoheres and error accumulates. The time a qubit can survive without losing its state information with high probability is called \textsf{coherence time}. The longer a circuit has to execute, the more likely it will approach a qubit's coherence time. IBM proposes the metric of \emph{quantum volume} \cite{Cross_2019} for evaluating the
effectiveness of quantum computers. One important factor for calculating quantum volume is the maximum depth of a circuit that can be executed by a quantum computer
before accumulating a certain amount of error. Here the depth represents the amount of time a circuit executes. Optimizing the depth of a circuit is important as only circuits that fit into the quantum volume can run successfully and generate meaningful computational results. Thus quantum compilers must take the execution time of the generated circuit into consideration.




. 



In this paper, we focus on time-aware qubit mapping. A good time-aware qubit mapper needs to yield a hardware-compliant circuit while having optimal or near-optimal execution time. We discover the key is to find intervals with slack in the circuit and to use the slack to hide the latency of inserted swap operations. We present important considerations for detecting and exploiting slack in the circuit. Our implemented qubit mapper named \textsf{SlackQ} automatically searches for dynamic qubit mappings given an input program on a quantum architecture with arbitrary qubit connectivity. The experiments show that SlackQ improves performance by up to 2.36X, by 1.62X on average, over 106 representative benchmarks from RevLib~\cite{Wille+:ISMVL08}, IBM Qiskit \cite{IBMQiskit}, and ScaffCC \cite{JavadiAbhari+:CCF14}.


\section{Background and Motivation}
\label{sec:motiv}
\subsection{Quantum Computing Basics}

\subsubsection{Qubit}
A quantum bit or qubit, is the counterpart to a classical bit in the realm of
quantum computing. Different from a classical bit that represents either `1' or
`0', a qubit is in the coherent superposition of both states. The state |$\psi$> associated with a qubit is a unit vector in a two-dimensional vector space. The state of a qubit can be represented as
\begin{gather*}
 |\psi> ~ = \alpha |0> + ~\beta |1> ~= 
  \begin{bmatrix}
   \alpha \\
   \beta 
   \end{bmatrix},
\end{gather*}
where $\alpha$ and $\beta$ are two complex numbers such that $|\alpha|^2 + |\beta|^2 = 1$. $\alpha$ and $\beta$ are called amplitudes. Upon the standard measurement, the state |$\psi$> will collapse into the basis state |0> with probability $|\alpha|^2$ or 
the basis state |1> with probability $|\beta|^2$. A system of n qubits encodes a state superposition of $2^n$ basis vectors with $2^n$ amplitudes. The classical n-bit system encodes the information of one basis vector in the vector space, but n-qubit system encode the information of $2^n$-dimensional vector space. Operating on one n-qubit state is as if operating on $2^n$ complex numbers at one time. This is one of the reasons for the potential  exponential speedup using quantum computing. 

\subsubsection{Quantum Gates}
There are two types of elementary quantum gates. One is the single-qubit gate, which is a unitary quantum operation that can be abstracted as the
rotation around the axis of the Bloch sphere \cite{Nielsen+:2002book}. 
A single-qubit gate can also be represented using a 2 by 2 unitary matrix. Important single-qubit gates include the H (Hadamard) gate, and the S (phase shift by $\pi/4$)
gate \cite{amy+:tcads13}. 

The second type of gate is the multi-qubit gate. The controlled-NOT (CNOT) is a two-qubit gate that performs the most important role (arguably) in quantum computation. The two qubits involved in a CNOT gate are: the control qubit and the target qubit. If the control qubit is 0, it leaves the target qubit unchanged. If it is 1, it applies a NOT gate to the target qubit. The CNOT gate entangles qubits and allow qubits to communicate. The CNOT gate, \textsf{H} gate, \textsf{S} gate, and \textsf{T} gate together form a universal set called the \textsf{Clifford+T} library. Any quantum algorithm can be implemented using a composition of gates from the universal set.

\subsubsection{Quantum Circuit}
A quantum algorithm can be expressed as a quantum circuit which is composed of a set of qubits and a sequence of quantum
operations on these qubits. A quantum circuit can be thought of a quantum algorithm in ``assembly language''.  There are two different ways to describe the quantum
circuits. One way is to use the quantum assembly language called
OpenQASM~\cite{Andrew+:arxiv17} released by IBM. The other way is to use a
circuit diagram, in which qubits are represented as horizontal lines. Input is the on the left and output is on the right. Unlike a classical circuit, a quantum circuit must have the same number of input and output qubits. Fig. \ref{fig:motivation}
(b) shows an example quantum circuit diagram. Logical qubits are denoted using lowercase letters (q1, q2, ...) and physical qubits are denoted using uppercase letters (Q1, Q2, ...). Initially, logical qubits q1, q2, q3, q4, and q5 are mapped to physical qubits Q1, Q2, Q3, Q4, and Q5. A single-qubit gate is denoted
as a square on the line. A CNOT gate is represented as a line connecting
two qubits where the control qubit is marked with a dot and the target qubit with a $\oplus$ sign. In this paper, we use the circuit diagram representation to describe examples.

\subsection{Qubit Mapping Problem}
To enable the execution of a quantum circuit, the logical qubits in the circuit must be mapped to the physical qubit on the target hardware. When applying a CNOT gate, the two logical qubits involved in
the CNOT gate must be mapped to two physical qubits connected to each other. Due to the irregular layout and connectivity of the qubits in the target device, it is sometimes impossible to find an initial mapping that makes the entire circuit CNOT-compliant. The common practice is to insert SWAP operations to remap the logical qubits, whenever a CNOT gate cannot be applied.

\begin{figure}[!htb]
  \centering
  \includegraphics[width=1.0\linewidth]{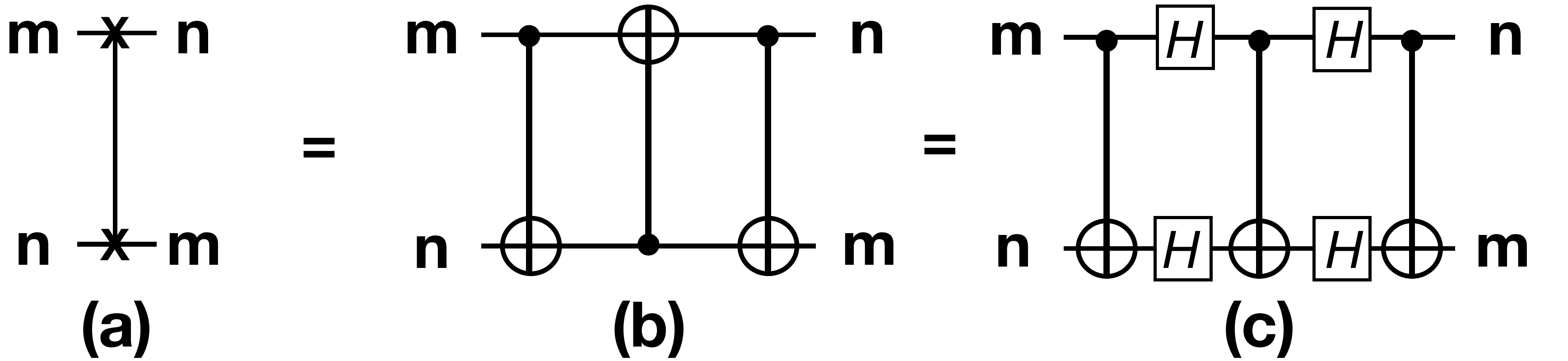}
  \caption{Implementation of a SWAP operation: (a) the SWAP notation, where m and n are two logical qubits, after SWAP, m and n exchanged their states, (b) for bidirectional links, where the three CNOT that implement the SWP do not need to use the same control qubit, and (c) for single direction links, where the three CNOT must use the same control qubit.}
  \label{fig:swapgate}
\end{figure}

A SWAP operation exchanges the states of the
two input qubits of interest. As shown in Fig. \ref{fig:swapgate} (b), a swap
operation is implemented using 3 CNOT gates for architectures with
bi-directional links, where a bi-directional link means both ends of the
link can be the control or target qubit. Or it can be implemented using 3 CNOT gates plus 4 Hadamard gates for architectures
with single-direction links as shown in Fig. \ref{fig:swapgate} (c), where a single-direction link means
only one end of the link can be the control qubit.

The qubit mapping problem takes a logical circuit and a hardware coupling graph as input and outputs a transformed circuit that fits on the hardware device by inserting SWAP operations. After the transformation, all CNOT gates must be performed on qubits that are connected in the physical architecture.  Due to the swaps, a logical qubit may be mapped to different physical qubits at different points in the circuit execution. But, at any given point, a logical qubit will be mapped to exactly one physical qubit since we are only using swaps to move qubits and are not making any copies. 
 
An example of circuit transformation is shown in Fig. \ref{fig:motivation} where (a) is the physical connectivity, (b) is the original circuit, (c) and (d) offer two different hardware-compliant circuits generated from the same original circuit.

\subsection{Parallelism in Quantum Circuit}
Like in classical computers, parallelism is also important in quantum computers. Parallelism comes from independent operations on different qubits. Gates on the same qubit have to run sequentially. For instance, if $a$ and $b$ are two consecutive gates on the same qubit, and $a$ is before $b$ in the program, then gate $b$ depends on $a$. Gates that do not share any qubit are independent. A two-qubit gate depend on up to two gates since it involves two qubits. A two-qubit gate has up to two immediate successors in the dependence graph. A dependence graph can be built with respect to the partial order between gates. It is a directed acyclic graph (DAG).


In a transformed circuit, the parallelism could be (1) between the gates in the original circuit (as \textsf{g2} and \textsf{g3} in Fig. \ref{fig:motivation} (b)), (2) between the SWAP gates that are inserted into the original circuit, and (3) between a gate in the original circuit and a newly inserted gate (as \textsf{swap 3,5} and g1 in Fig. \ref{fig:motivation}(c)). A good qubit mapping algorithm should consider all types of parallelism. However, existing studies only consider the first two types of parallelism. Our work is the first one that systematically exploits all type of parallelism. 

As shown in Fig. \ref{fig:motivation}, the best two known approaches by Zulehner \etal
\cite{Zulehner+:DATE18} and Li \etal \cite{Li+:ASPLOS19} do not distinguish solution (c) from solution (d) as the two solutions both 
insert 1 SWAP. And \cite{Zulehner+:DATE18} and \cite{Li+:ASPLOS19} only optimize the number of gates inserted into the
circuit (or the parallelism of the inserted gates), but not the parallelism of the
transformed circuit. The solution in Fig. \ref{fig:motivation} (c) is better than Fig.  \ref{fig:motivation} (d) as the inserted swap can run in parallel with the gates in the original circuit. This example stresses the importance of
time-awareness in SWAP insertion schemes and motivates our work.

\section{Insight and Design}
\label{sec:insight}
To improve the parallelism between the inserted SWAP operation and the gates in the original circuit, we discover that it is important to exploit the \textsf{slack} intervals in the circuit. The slack represents the idle time in the original circuit for a given set of qubits. The key is to hide the latency of inserted swap operations by using the qubits that are idle at that time of the circuit execution. This forms the main idea of this paper and we insert SWAP operations such that they leverage \textsf{slack} in the circuit as much as possible.

\subsection{Slack}
\label{sec:slack}
We define \textsf{slack} as the idle time between two consecutive gates on the same qubit(s) and can be used to perform SWAP operation without affecting the total execution time of the entire circuit. The \textsf{slack} time is usually caused by dependence between gates and/or variation of gate count on individual qubits.

\begin{figure}[h]
  \centering
  \includegraphics[width=0.75\linewidth]{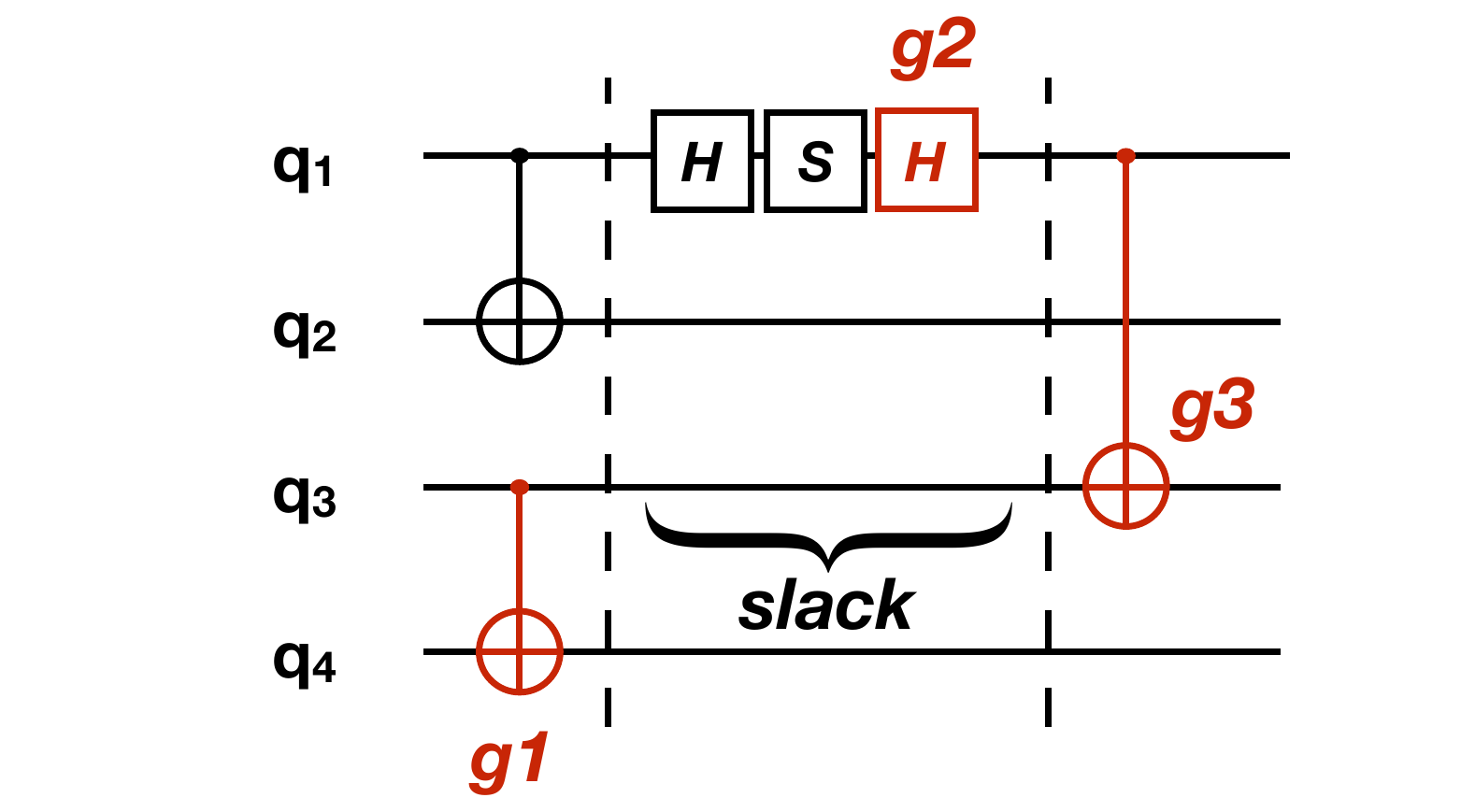}
  \caption{Slack in the circuit: Since \textsf{g3} depends on \textsf{g1} and \textsf{g2}, \textsf{g1} finishes earlier than g2, therefore, for qubit \textsf{q3}, there is a slack interval of three cycles (assuming each gate takes one cycle) between \textsf{g1} and \textsf{g3} in the circuit. }
  \label{fig:slack_cycles}
\end{figure}

The \textsf{slack} time due to dependence between gates only occurs when there are two-qubit gates in the circuit. Recall that a CNOT gate depends on up to two other gates, since CNOT is a two-qubit gate. If the qubits are running at different speeds, one of the other qubits might be ready earlier than the other. The faster qubit thus needs to wait for the slow qubit to finish before the CNOT gate can be executed. On the other hand, if a circuit has a number of qubits, and the number of gates on each qubit is different (even if they are all independent), then some qubits will inevitably be idle at some point of the execution. The slack intervals can be used for inserting swap operations that resolve qubit mapping constraints. An example of \textsf{slack} in the circuit is shown in Fig. \ref{fig:slack_cycles}. 


\begin{figure*}[htb]
  \centering
  \includegraphics[width=0.9\linewidth]{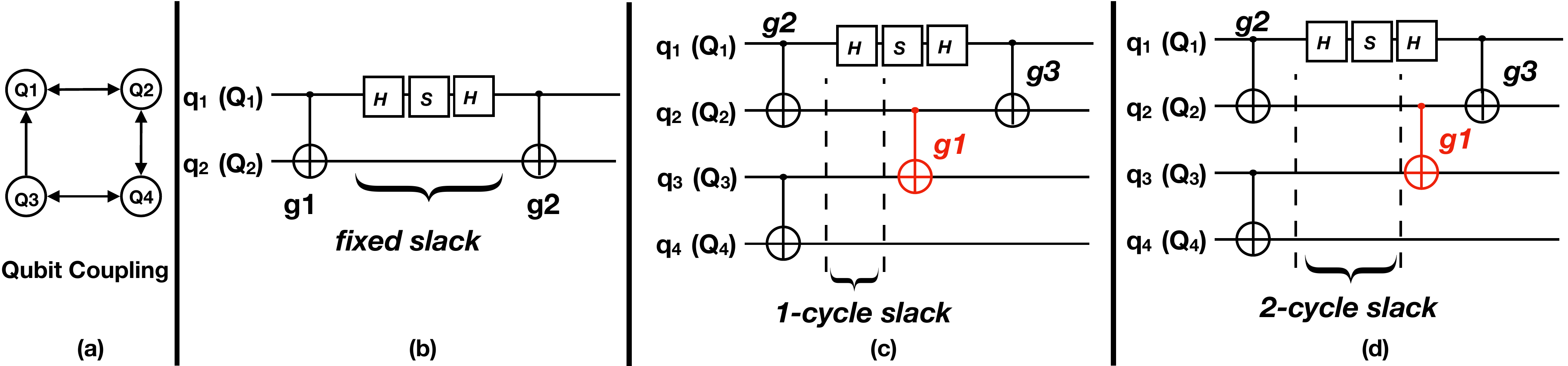}
  \caption{(a) Qubit coupling graph; (b) An example of fixed slack in the quantum circuit; (c) and (d) Examples of flexible slacks, where \textsf{g1} can be moved within a time window without affecting the circuit execution time, the slack before g1 can be either 1 cycle or 2 cycles assuming every gate takes one cycle. Note the slack between \textsf{g1} and \textsf{g3} may also vary due to scheduling of g1. }
  \label{fig:remainingcircuit}
\end{figure*} 

There are two types of slacks in the circuit. One type does not require the rescheduling of the gates, and we define it as \textsf{fixed slack}. The other type of slacks may have variable number of cycles, and we denote it as \textsf{flexible slack}. A good qubit mapper needs to search globally and exploit both \textsf{fixed} and \textsf{flexible slack}.

\vspace{-5pt}
\paragraph{\textbf{Fixed Slack}} An example of \textsf{fixed slack} is shown in Fig. \ref{fig:remainingcircuit} (b). Assuming each gate takes one cycle, there is a fixed slack between \textsf{g1} and \textsf{g2} on qubit \textsf{q2}. Here it cannot delay g2 or start g1 early if the total execution time needs to remain unchanged. If qubit \textsf{q2} is used to perform another gate such as the swap operation during the three cycles, it will not affect the execution time of the entire circuit. In this case, the number of cycles that can be used on \textsf{q2} between g1 and g2 is fixed. 

\begin{figure}[!ht]
  \centering
  \includegraphics[width=1.0\linewidth]{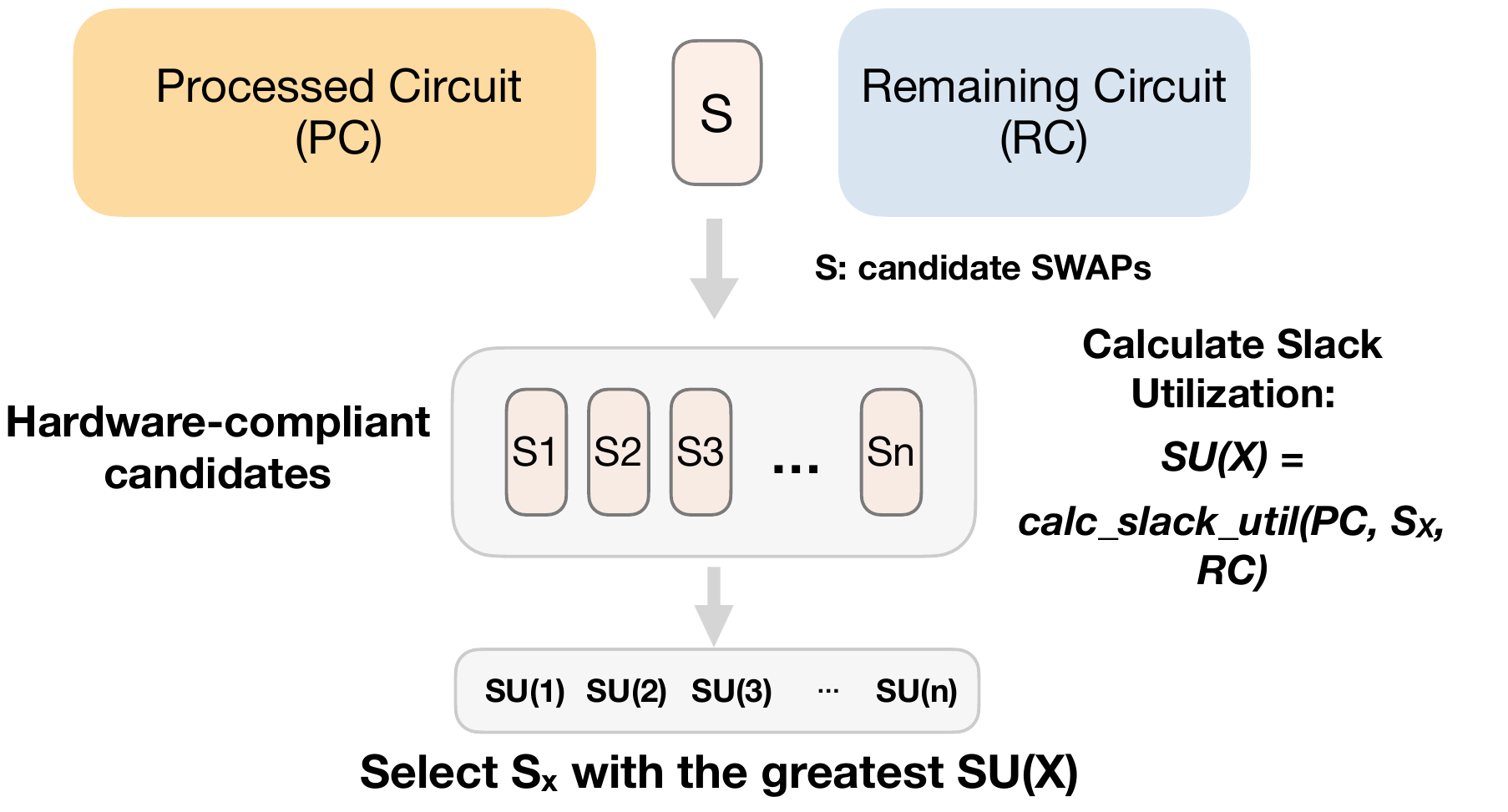}
  \caption{Choose SWAP Candidates}
  \label{fig:algo}
\end{figure}

\begin{figure}
    \centering
    \includegraphics[width=0.5\textwidth]{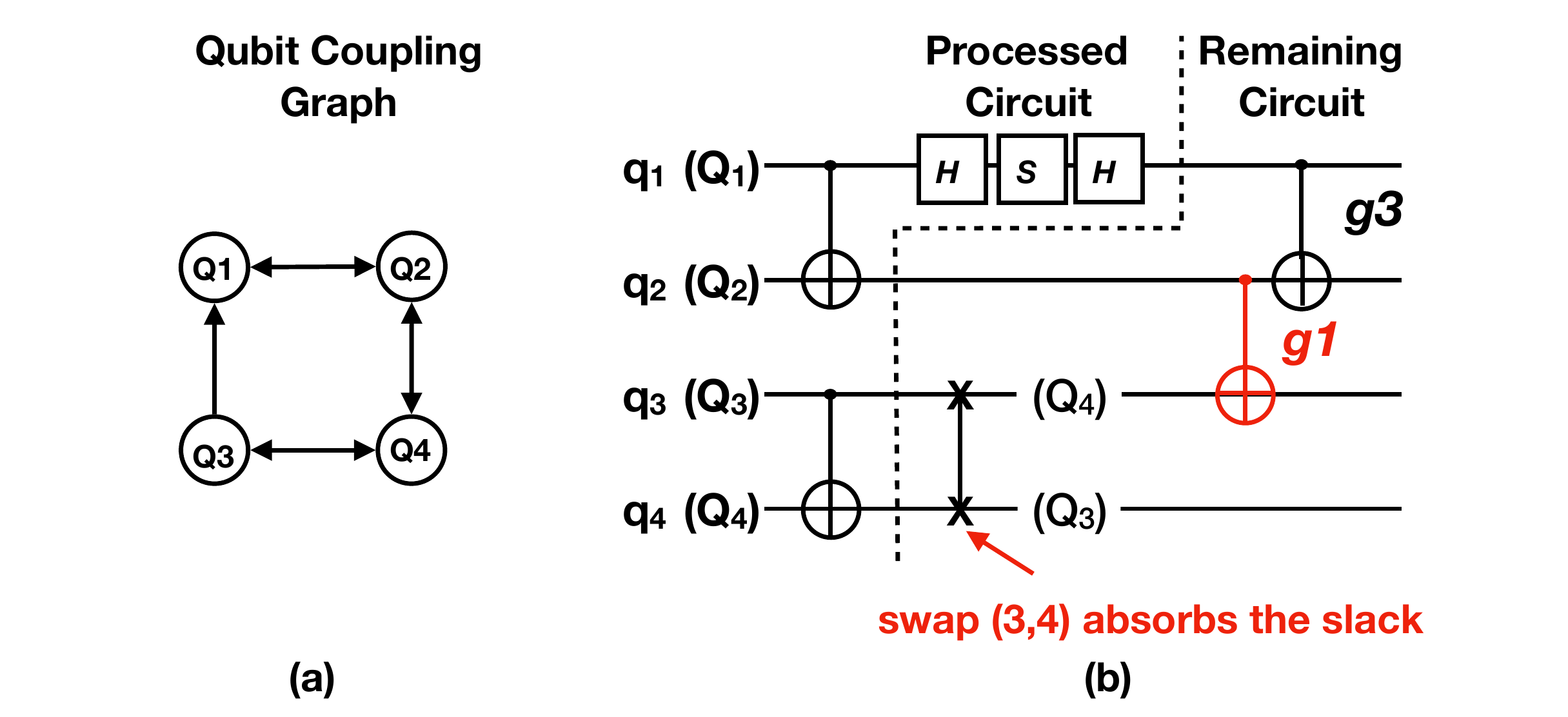}
    \caption{Scheduling gates to create more slacks: gate \textsf{g1} can be moved forward such that \textsf{swap Q3, Q4} can absorb the longer slack on q3}
    \label{fig:flexibleslack}
\end{figure}

\vspace{-10pt}
\paragraph{\textbf{Flexible Slack}} Sometimes fixed slack does not always exist. It is necessary to move the gates in order to create slacks for latency hiding purpose. We show an example Fig. \ref{fig:remainingcircuit} (c) and (d), where slack can be created by moving $g1$. Let's say cnot($q{1}, q{2}$) and cnot($q_{3}, q_{4}$) are scheduled on cycle 1. The three single-qubit gates on $Q_{1}$ are scheduled on cycle 2,3,4 respectively. With this going on, $g{3}$ expects to be executed on cycle 5 at the earliest. $g{3}$ depends on $g{1}$. $g{1}$ can be scheduled at the second cycle, the third cycle (Fig. \ref{fig:remainingcircuit} (c) ) or the fourth cycle (Fig. \ref{fig:remainingcircuit}  (d)) without delaying $g5$. To this end, a slack with zero, one or two cycles can be created between $g2$ and $g1$, depending on when $g1$ is scheduled. And this type of slack between $g2$ and $g1$ is flexible. On the other hand, since $g1$ is not directly executable due to the connectivity constraint in Fig. \ref{fig:remainingcircuit} (a), the more slack intervals before $g1$ there are, the better it is for hiding the swap latency. In Fig. \ref{fig:flexibleslack}, we show that by moving $g{1}$ forward, q2 and q3 can have more slack intervals before $g1$, and swap(3,4) is inserted which utilizes the slack, resulting in a total circuit time of 6 cycles only, which is optimal in this case. 

It is worth mentioning flexible slack could be cascading as the rescheduling of one gate might affect its descendants or predecessors. For the fixed slack, the gates involved cannot be delayed without affecting the circuit time. Flexible slack allows one or multiple gates to delay start within reasonable time window(s). Flexible slack are more complicated than fixed slack. It is necessary to analyze and exploit flexible slack in a systematic way.

\vspace{-5pt}
\subsection{Dynamic Gate Scheduler}
We model the resolution of qubit mapping conflicts as a dynamic scheduling process. Gates in the circuit are scheduled as soon as their dependencies are resolved. When a gate cannot be scheduled due to a connectivity problem, we insert a (combination of) swap(s) to change the qubit mapping so that the gate can be executed on the physical device. All the gates that have already been scheduled at one point of scheduling are called the \textsf{Processed Circuit}, and the gates that still await scheduling are called the \textsf{Remaining Circuit}. 

\begin{figure*}[!ht]
  \centering
  \includegraphics[width=1\linewidth]{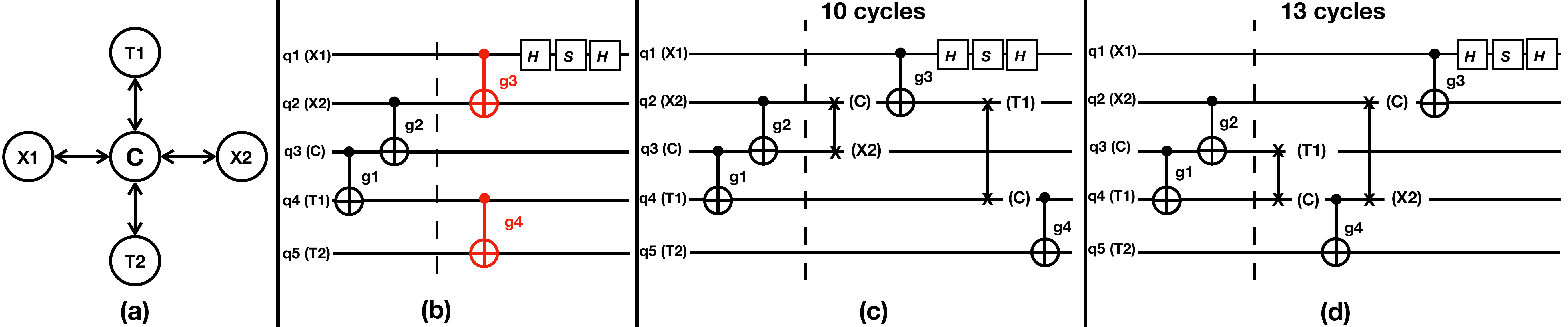}
  \caption{(a) Qubit Connectivity for a five-qubit machine (b) Original Circuit before qubit mapping (c) \textbf{Strategy One:} Resolving gates on critical path first (d) \textbf{Strategy Two:} Resolving gates on non-critical path first. }
  \label{fig:criticality}
\end{figure*}

Fig. \ref{fig:flexibleslack} shows an example of how the scheduling works. With initial mapping of $\{q_{1}, q_{2}, q_{3}, q_{4}\} \to \{Q_{1}, Q_{2}, Q_{3}, Q_{4}\}$, the first two CNOTs \textsf{cnot($q_{1}, q_{2}$)} and \textsf{cnot($q_{3}, q_{4}$)} and the three single-qubit gates on $Q{1}$ can be scheduled without remapping. At this point, those gates that are scheduled are part of the \textsf{Processed Circuit}. The remaining two CNOT gates ($g{1}$ and $g3$) that cannot be scheduled are part of the \textsf{Remaining Circuit}. Gate $g{1}$ cannot be scheduled because Q2 and Q3 are not connected in the device. Gate $g{3}$ cannot be scheduled because $g{1}$ must be scheduled before $g{3}$ (write-after-read dependency on Q2). The dashed lines divide the circuit into processed part and remaining part. 

To minimize the circuit time, we search for swap candidates for $g{1}$ that results in maximally hiding swap latencies using circuit slack. Fig. \ref{fig:algo} shows the key idea behind the searching for optimal swap candidates. The search reveals multiple hardware-compliant candidates that utilize different sequences of swaps to achieve compliance.  We choose the optimal candidate by calculating the \emph{Slack Utilization} of each candidate, and choosing the one with the best utilization. In Fig. \ref{fig:flexibleslack}, we choose swap candidate \textsf{swap(q3, q4)} since it best hides the swap latency behind the 2-cycle slack shown in Fig. \ref{fig:remainingcircuit} (d). Now $g{1}$ is satisfied and scheduling can proceed. 


\subsection{Critical Gates}
The gates in the remaining circuit  pending scheduling whose dependences have been resolved but connectivity problems haven't been can be divided into two groups: those on the critical path and those that are not. We denote the gates on the critical path as \textsf{critical gates}, and the others as \textsf{non-critical gates}. In parallel computing, the critical path length is equal to the execution time when there is enough parallelism. In this case, the critical path is equal to execution time as the maximum parallelism (the maximum number of gates that can run concurrently) is at most the same as the number of qubits. Thus it is important to prioritize the scheduling of critical gates over non-critical gates. 

To prioritize critical gates, what we need to do is to resolve the connectivity problems of critical gates as early as possible. Imagine a scenario where two gates have connectivity problems, one gate is critical and the other is non-critical gate. Their connectivity issues cannot be resolved at the same time. Under this situation, we should resolve the critical gate first, as resolving the \textsf{non-critical} gates can be likely delayed without affecting the overall execution time.

We use an example from Fig. \ref{fig:criticality} to show how \textsf{criticality} can play an important role in determining the overall circuit time. We use a five-qubit quantum machine, whose connectivity is shown in Fig. \ref{fig:criticality} (a). This example circuit consists of 4 CNOTs and 3 single-qubit gates, with $g{1}$, $g{2}$ scheduled, and $g{3}$, $g{4}$ not yet scheduled due to connectivity issues. It's crucial to note that $g_{3}$ is on the critical path, while $g{4}$ is not. The two gates $g{3}$ and $g{4}$ cannot be resolved at the same time if both of them want to use only one swap, since qubit $C$ is the on the path from $T1$ to $T2$, and from $X1$ to $X2$. Whether to prioritize $g{3}$ over $g{4}$ when using the hub qubit $C$ for swap, makes a big difference in terms of circuit time. We show this discrepancy by illustrating two strategies and their resulting circuits. 
\begin{itemize}
\item {\textbf{Strategy One - Prioritizing critical gates}}  Resolve $g{3}$ first. Shown in Fig \ref{fig:criticality} (c), it is necessary to insert swap($q2$, $q3$) before $g{3}$. After $g{3}$ is resolved and scheduled, swap($q2$, $q4$) is inserted such that $g{4}$ can be resolved. swap($q2$, $q4$) can take advantage of the slack on logical qubits $q2$ and $q4$, as logical qubit $q1$ is processing three single-qubit gates. This strategy results in total circuit time of 10 cycles, assuming CNOT and single-qubit gate both have latencies of one cycle.

\item {\textbf{Strategy Two - Not distinguishing  critical gates from non-critical path}} Resolve $g{4}$ first. Shown in Fig \ref{fig:criticality} (d), it is necessary to insert swap($q3$, $q4$) before $g{4}$, and let $g{4}$ be scheduled. In the meantime, $g{3}$ has to wait, which results in the critical path being elongated due to the delay of the execution time of $g{4}$ and swap($q3$, $q4$).  It is because when $g{4}$ is being executed, the mapping that allows $g{4}$ must be kept, which will delay all the remaining gates. In this case, it is not desirable to delay all the remaining gates as they are on critical path. Delaying gates that are  critical will have a more detrimental impact than delaying gates not on critical path. After $g{4}$ is resolved, the fastest way to resolve $g{3}$ is to swap($q2$, $q4$) before $g{3}$. This strategy as a whole results in total circuit time of 13 cycles, which is 30\% more than strategy one. 
\end{itemize}

It can be seen from this example the later resolving of the non-critical gates are highly likely to overlap with the gates on the critical path, and result in less impact to overall circuit execution.

\section{Implementation}

\label{sec:implemeatation}
Based on the design consideration on Section \ref{sec:insight}, we implement a slack-aware qubit mapping framework called \textsf{SlackQ}. 

\subsection{Overview of \textsf{SlackQ} }
Our algorithm is an iterative gate scheduler which dynamically resolves the connectivity issues encountered during the scheduling process. Initially, a dependency graph of the circuit is built. Then we traverse the dependency graph of the circuit and schedule the gates one by one. We keep a frontier set of gates ready to be scheduled.   
 When resolving the connectivity issues, we invoke a priority-queue based searcher for swap candidates. It returns hardware-compliant candidates.  
 Among these hardware-compliant candidates, the one that has the best slack utilization is  chosen, and the scheduling process proceeds.
 We describe the algorithm below with respect to the pseudo-code shown in Algorithm \ref{algo:search}:

\noindent \textbf{Step One - Initialization} This step prepares for the searching process. It builds the dependency graph of circuit. It finds the gates that do not depend on any other gates. Then it places those gates into the frontier $F$. It also initializes the processed gate set $P$ as empty set, and the remaining gate sets $R$ as the entire circuit. 

\noindent {\textbf{Step Two - Schedule Ready Gates}} This step goes through frontier list $F$. It finds all gates in $F$ that can be scheduled immediately due to having no connectivity issues according to the current mapping $\pi$. It schedules all these gates. When finishing the scheduling of one gate, it finds the descendant gate and see if this  descendant's other parent has also been scheduled. If this is the case, the descendant gate's dependency is resolved. It then places this descendant gate into $F$. This step is repeated until $F$ contains no gate that can be scheduled with respect to the current qubit mapping.

\noindent {\textbf{Step Three - Resolve Qubit Mapping Conflicts}} We go through the frontier $F$ again, finding the gates with resolved dependencies but are constrained by the current mapping and are on the critical path of the remaining circuit. Put those gates into a set called $F_{critical}$. Run a priority queue based searcher for hardware-compliant mappings. Our mapping searcher here returns a list of hardware-compliant mappings candidates, called $M$. 
Among these candidates, it finds the one (call it $m$) with the best slack utilization. Then we use the swap sequence associated with $m$ to update the mapping $\pi$ and add the swap sequence into the processed circuit P. 

\noindent {\textbf{Step Four}} Repeat Step Two and Three until all gates are scheduled in the circuit. Return transformed circuit. 

In Sections \ref{sec:impl_init} to \ref{sec:impl_resolve_conf}, we describe a few important aspects of this algorithm.
\vspace{-10pt}
\subsection{Initialization}
\label{sec:impl_init}
Before calling the scheduler described in Algorithm \ref{algo:search}, we initialize the frontier $F$ and processed circuit P, and the remaining circuit R. Initially, P is empty and R is the entire circuit. For $F$, it creates a Directed Acyclic Graph~(DAG) to represent the dependency between quantum operations.  Fig. \ref{fig:update_graph} (a) shows an example of a dependency graph from the circuit illustrated in Fig. \ref{fig:motivation} (b).  



{\small
\begin{algorithm}[htb!]
\SetAlgoLined
\SetKwInOut{Input}{Input}
\SetKwInOut{Output}{Output}
\Input{Frontier $F$, initial mapping $\pi$, processed circuit $P$, remaining circuit $R$}
\Output{Transformed circuit T}
\BlankLine

\While{F not empty}{
    E = getSchedulableGates(F, $\pi$)\;
    \While{E not empty}{
        F.remove(E)\;
        P.add(E)\;
        R.remove(E)\;
        \For{g $\in$ E}{
            \For{d $\in$ g.children}{
                \If{d's dependency is resolved}{
                    F.add(d)\;
                }
            }
        }
        E = getSchedulableGates(F, $\pi$)\;
    }
    $F_{critical}$ = select\_critical\_gates ($F$)\;
    mapping\_candidates = resolve\_conflicts ($F_{critical}$, $\pi$)\;
    m = best\_slack\_utilization (mapping\_candidates, P, R)\;
    $\pi$  = update\_mapping\_with\_swaps(m, $\pi$)\;
    P.add(m.swaps)\;
}
return P\;
\caption{Dynamic Gate Scheduler} 
\label{algo:search}
\end{algorithm}
}

\vspace{-7pt}
\subsection{Choosing the Best Mapping Candidate}
With multiple hardware-compliant mappings, it is necessary to determine the candidate that has the best slack utilization. The best slack utilization means the inserted swap sequence makes best use of the slack currently existing in the circuit.
To evaluate these mapping candidates, for each of them, we tentatively insert the associated swap sequence and monitor how the swap insertions affect the dependence graph and the critical path. We trace the nodes that are affected due to the inserted swaps and detect how much their start/ending time changes. 

\begin{figure*}[!ht]
  \centering
  \includegraphics[width=0.95\linewidth]{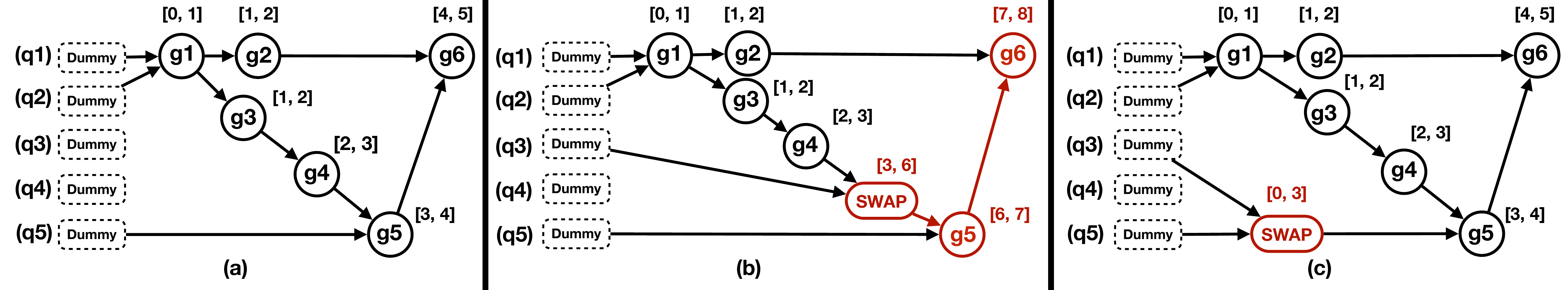}
  \caption{(a) The generated dependency graph from the example of Fig. \ref{fig:motivation}. The numbers displayed on each gate refer to the start/end cycle of this gate.   (b) One mapping candidate whose inserted swap results in two later gates delaying its start/end cycles. (C) Another mapping candidate whose inserted swap does not affect the start/end cycles of the entire circuit.}
  \label{fig:update_graph}
\end{figure*} 

We again use the example circuit and qubit coupling graph in Fig. \ref{fig:motivation} to show how our evaluation approach works. We first show the original dependency graph in Fig. \ref{fig:update_graph} (a). The numbers on gates denote the start/end cycle of each gate. For instance, \textsf{[0, 1]} represents the start cycle as 0 and the ending cycle as 1. We assign a dummy gate node at the beginning of the circuit for each qubit $q1$ \char`\~ ~$q5$, for the sake of illustration. The dummy node starts at time 0 and takes 0 cycles. In this example, we have two possible mapping candidates whose swap sequences are to be inserted on different qubits. We need to choose the better one out of the two mapping candidates. The first mapping candidate shown in Fig. \ref{fig:update_graph} (b) inserts one swap on logical qubits $q2$ (the qubit for $g4$) and  $q3$ in between gates $g4$ and $g5$, corresponding to the circuit in Fig. \ref{fig:motivation} (d). It affects gates $g5$ and $g6$, which are marked in red. For each affected node in the dependence graph, to calculate its earliest start time, one needs to check each of its parent nodes' ending time, choose the maximum one, and use it as its own start time. In this example, added swap result in change in g5's start time as well as the change in g6'start time, and delays the entire circuit by 3 cycles. Here we assume each gate takes one cycle. We assume a swap is implemented using 3 CNOT gates and thus is 3 cycles. 

The second mapping candidate shown in Fig. \ref{fig:update_graph} (c) inserts one swap on logical qubits $q3$ and $q5$ placed right in front of $g5$. It results in no changes to the start/end cycles of the entire circuit, since there is slack on physical qubits $q3$ and $q5$. Obviously, it should choose the mapping candidate illustrated in Fig. \ref{fig:update_graph} (c). 

We use an algorithm to systematically analyze the start/ending time of each gate due to inserted swaps. The algorithm does not have to traverse the entire dependence graph. Instead, it only traces the affected gates in the original circuit to find the candidate that leads to the smallest increment to total circuit time. We also add an optimization to quickly terminate the tracing when a candidate is deemed hopeless. A candidate is deemed hopeless if one of the affected gates is on the critical path and the delay to that gate due to swaps already exceeds the smallest increment found in a previous candidate. In that case, it terminates the tracing and moves on to the next candidate. 
The algorithm is described in Algorithm \ref{algo:update_graph}.

{\small
\begin{algorithm}[htb!]
\SetAlgoLined
\SetKwInOut{Input}{Input}
\SetKwInOut{Output}{Output}
\Input{Mapping candidates $M$, dependency graph $G$, proessed circuit $P$}
\Output{Best slack-utilizing mapping $m_{best}$}
smallest\_inc = $\infty$\;
CP = getCriticalPath(G)\;
\For{$m \in M $}{
    $RG$ = getLastScheduledGateOnEachQubit(P)\;
    G' = G\;
    G'.addGates(m.swaps)\;
    graph\_updated = True\;
    circuit\_time = 0\;
    \While{$RG$ not empty}{
        $RG'$ = []\;
        \For{$g \in RG$}{
            g.updateTentativeStartAndEndCycle()\;
            delta = g.tentativeStart - g.originalStart\;
            \If{g is on critical path \& delta > smallest\_inc}{
                graph\_updated = False\;
                break the while loop\;
                
            }
            \If{delta > 0}{
                RG'.add(g.children)\;
            }
            circuit\_time = max(circuit\_time, g.tentativeEnd)\;
        }
        RG = RG'\;
    }
    \If{(graph\_updated == True \& circuit\_time $>$ CP)}{
        \If{ ( (circuit\_time - CP) $<$ smallest\_inc )} {
        smallest\_inc = circuit\_time - CP \;
        $m_{best}$ = m\;}
    }
}
return $m_{best}$\;

\caption{Find the best slack-utilizing mapping} 
\label{algo:update_graph}
\end{algorithm}
}

\vspace{-10pt}
\subsection{Navigating the Candidate Search Space}
\label{sec:impl_resolve_conf}
We use a priority queue based searcher for qubit mapping candidates. The search space consists of state nodes that represent possible mappings from logical qubits to physical qubits. A mapping can be represented as $\pi: \{q_{1}, q{2}, ..., q_{n}\} \to  \{Q_{1}, Q{2}, ..., Q_{n}\}$. Applying swaps on top of a mapping can convert it into another mapping. Specifically, if we apply "$swap$ $q_{i}, q_{j}$" on a certain mapping $\pi_{old}$  and create the resulting mapping $\pi_{new}$, we will have $\pi_{new}[q_{i}] = \pi_{old}[q_{j}]$ and $\pi_{new}[q_{j}] = \pi_{old}[q_{i}]$.

Given $F_{critical}$ and current mapping $\pi$, it starts searching the state space of all feasible mappings that satisfy $F_{critical}$. It picks a node to expand and enumerate all possible parallel one-step swaps as the node's successors. We use a priority queue that is similar to that in \cite{Zulehner+:DATE18}. Unlike the work by \cite{Zulehner+:DATE18} where the search stops when the first state node that resolves all connectivity conflicts is retrieve from the priority queue, our search stops after $m$ expansions since the mapping candidate with minimal swap count is found, or when the gate count of the mapping candidate that is just retrieve has less than or equal to $k$ times more gates than the minimal swap count. We set $m=20$ and $k=2$ such that the returned mapping candidate will have reasonable gate counts. After all mapping candidates have been retrieved, we rank them with respect to the metric of best slack utilization discussed above.

\section{Evaluation}
\label{sec:eval}
In this section, we evaluate our  slack-aware swap insertion scheme (\textbf{SlackQ}) and compare it with the two state-of-the-art qubit mappers, respectively by \cite{Zulehner+:DATE18} and \cite{Li+:ASPLOS19}.  

\subsection{Experiment Setup}

    {\textbf{Benchmark.}} We use 106 benchmarks from RevLib~\cite{Wille+:ISMVL08}, IBM Qiskit \cite{IBMQiskit}, and ScaffCC \cite{JavadiAbhari+:CCF14}. RevLib comprises of a collection of benchmarks in the domain of reversible and quantum circuit design. Qiskit is a programming framework for quantum computing provided by IBM. ScaffCC is a compilation framework for the Scaffold quantum programming language. These benchmarks feature functionalities from implementing ALU logics, comparing inputs with constant values, ternary counters, to classic quantum algorithms like Quantum Fourier Transform (QFT) and ising model. 
    
     \noindent  {\textbf{Baseline}} We compare our work with two best known qubit
mapping solutions ~\cite{Zulehner+:DATE18}
(denoted as \emph{Zulehner}) and the Sabre qubit mapper from~\cite{Li+:ASPLOS19}
(denoted as \emph{Sabre}). We also compare our results with IBM's stochastic mapper in Qiskit \cite{IBMQiskit}. Since IBM's Qiskit mapper is significantly
worse in terms of circuit time than all other mappers we have evaluated, we do not show the results. The performance of Qiskit mapper is also noted in the work by \cite{Zulehner+:DATE18}.

    \noindent {\textbf{Metrics}} We compare the execution time of the transformed circuits generated by different qubit mapping strategies. It is worth mentioning that our approach can take any gate latency as input parameters and generate transformed circuits based on the input. However, to make evaluation results as close to real machines as possible, we use the results from the studies by \cite{murali+:asplos19, linke+:nas2017}. In these studies, different types of quantum architecture are investigated, and the studies reveal that two-qubit gates usually takes around twice as much time as single-qubit gates. Hence we assume single-qubit gates take 1 cycle and two-qubit CNOT gates take 2 cycles in our experiments. The time is reported as the total number of executed cycles.

  \noindent {\textbf{Platform}} We use IBM's 20-qubit Q20 Tokyo architecture \cite{Li+:ASPLOS19} as the underlying quantum hardware. The qubit mapping approach is implemented in C++ and  executed on a Intel 2.4 GHz Core i5 machine, with 8 GB 1600 MHz DDR3 memory. 

\subsection{Experiment Analysis}

\begin{figure}[!ht]
  \centering
  \includegraphics[width=0.8\linewidth]{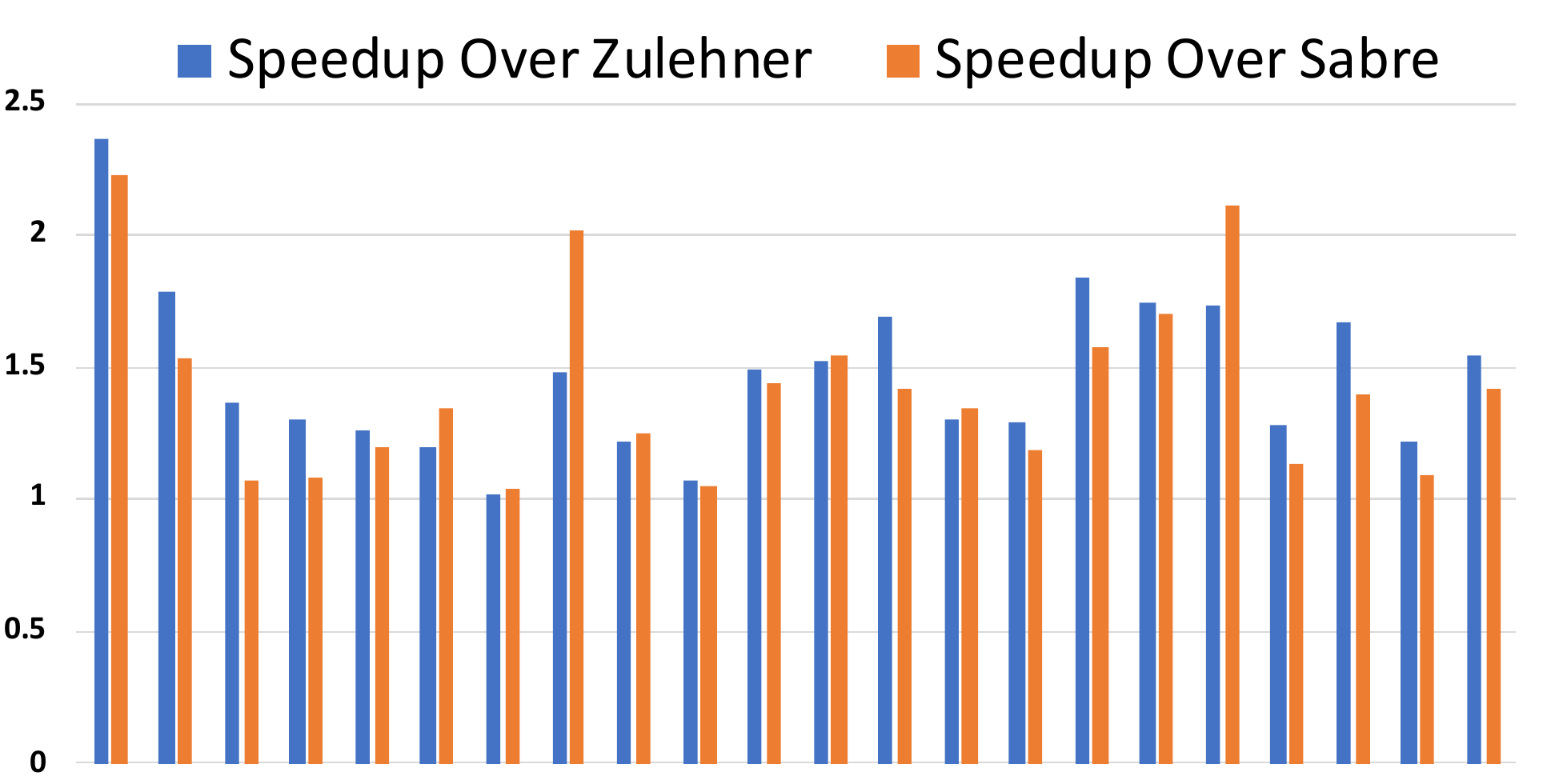}
  \caption{Speedup for Mini Benchmarks (< 200 gates)}
  \label{fig:mini_bench}
\end{figure}

\begin{figure}[!ht]
  \centering
  \includegraphics[width=0.8\linewidth]{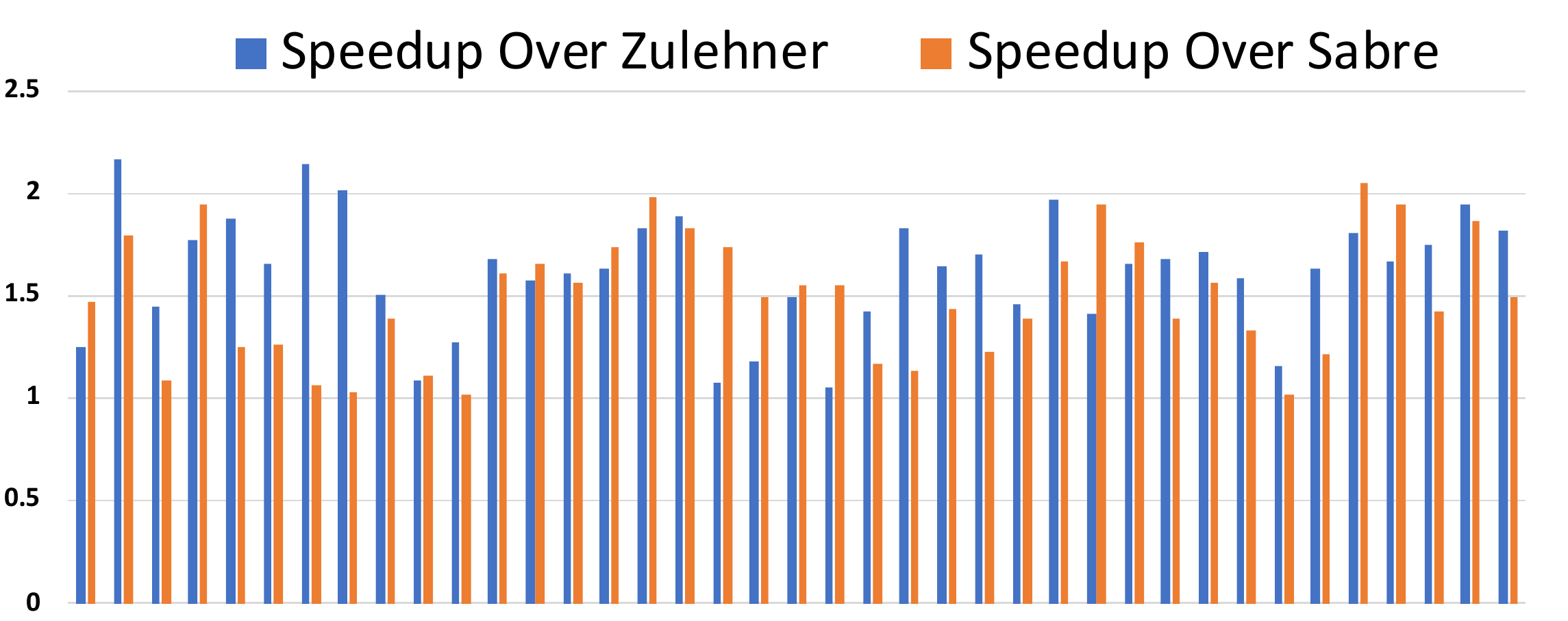}
  \caption{Speedup for Small Benchmarks (< 1000 gates)}
  \label{fig:small_bench}
\end{figure}

\begin{figure}[!ht]
  \centering
  \includegraphics[width=0.8\linewidth]{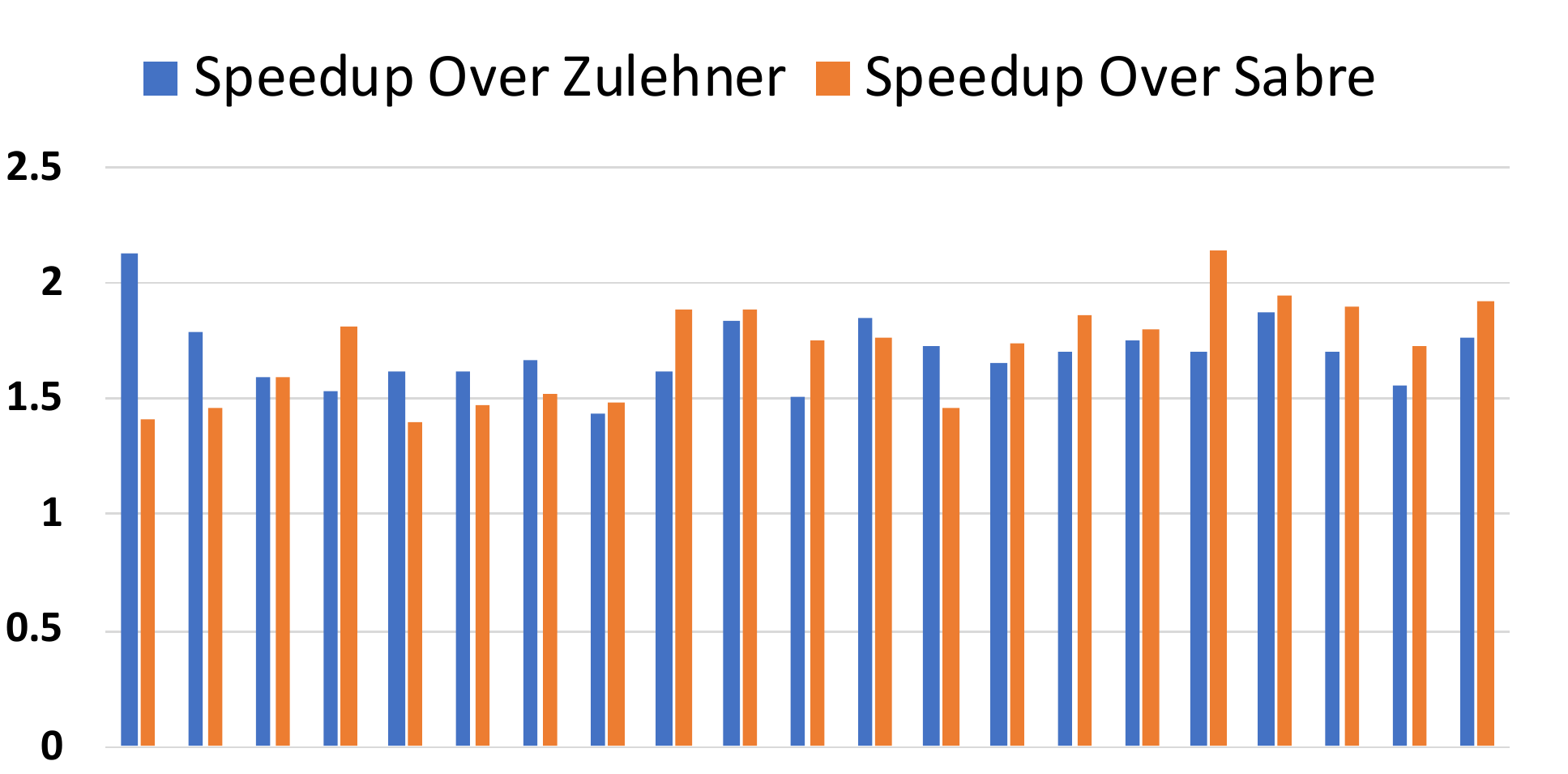}
  \caption{Speedup for Medium Benchmarks (< 10000 gates)}
  \label{fig:medium_bench}
\end{figure}

\begin{figure}[!ht]
  \centering
  \includegraphics[width=0.8\linewidth]{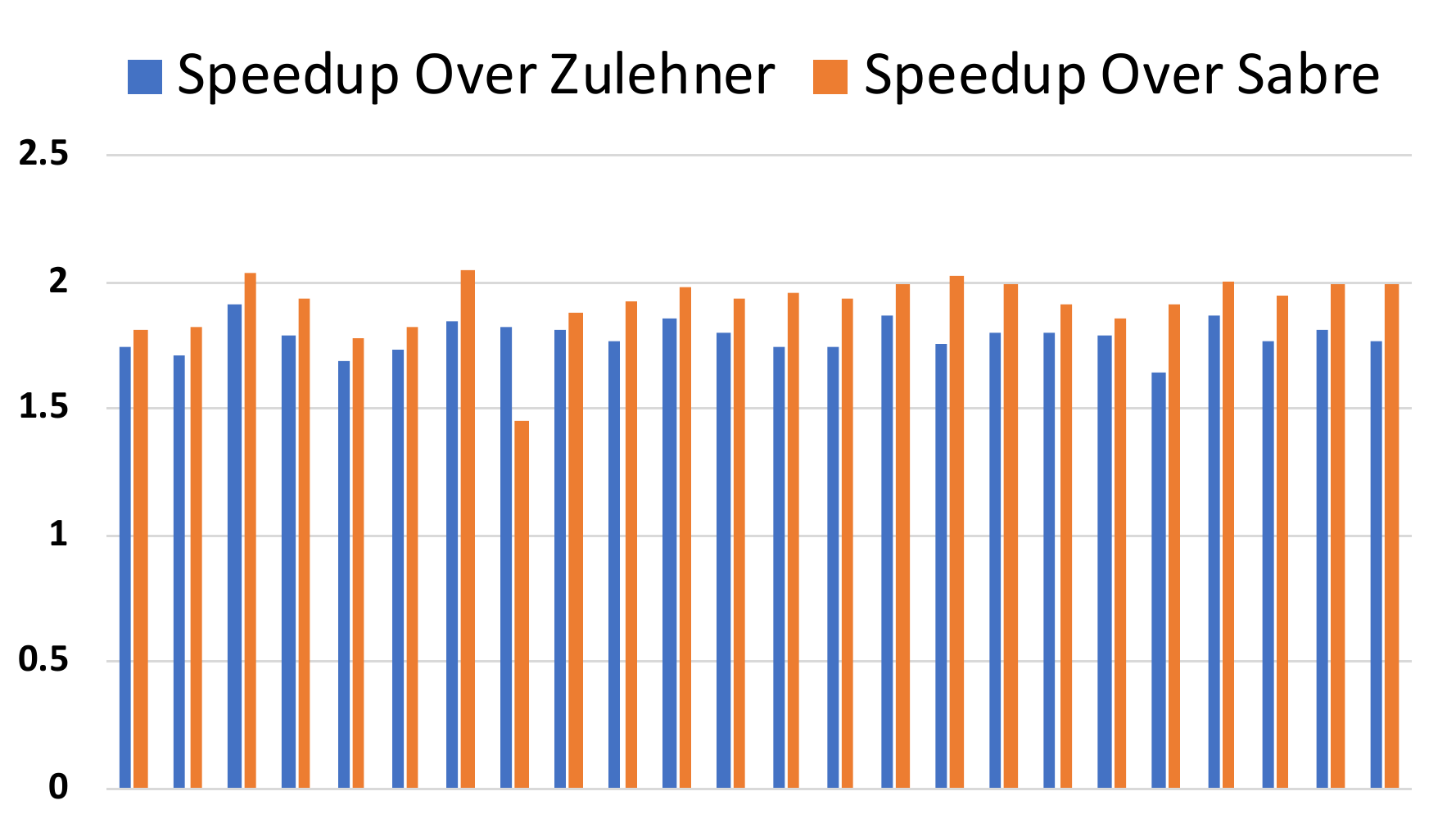}
  \caption{Speedup for Large Benchmarks (< 200000 gates)}
  \label{fig:large_bench}
\end{figure}

We categorize the 106 benchmarks into four categories. Benchmarks in the first category each has less than 200 gates, and we denote them as \textsf{mini} benchmarks. There are 22 mini benchmarks. The second category has benchmarks with 200 to 1,000 gates. We name this category as \textsf{small} benchmarks. There are 39 small benchmarks. The third category of benchmarks have 1,000 to 10,000 gates. We name it as \textsf{medium} benchmarks and there are 21 benchmarks in this category. The fourth category of benchmarks have 10,000 to 200,000 gates. We refer to it as \textsf{large} benchmarks and there are 24 benchmarks in this category. The results for \textsf{mini}, \textsf{small}, \textsf{medium}, and \textsf{large} benchmarks are presented in Fig. \ref{fig:mini_bench}, Fig. \ref{fig:small_bench}, Fig. \ref{fig:medium_bench}, and Fig. \ref{fig:large_bench} respectively. 

It can be observed from the results that as the problem size scales, the performance improvement brought by \textsf{SlackQ} improves. For most benchmarks in the mini and small category, the speedup is between 1.1X and 1.5X. However, for the medium and large category, the speedup for most benchmarks is above or around 1.5X. The average speedup for mini benchmarks is 1.45X and for small, medium, and large benchmarks, the average speedup becomes 1.55X, 1.70X, and 1.86X respectively. The results show that our approach works well in general, and in particular for larger benchmarks.

There are two baselines we compare against: the Zulehner approach and the Sabre approach. For mini and small benchmarks, the \textsf{Zulehner} approach does not seem to perform as well as the \textsf{Sabre} approach. It can be seen from the fact that the relative speedup of \textsf{SlackQ} over \textsf{Zulehner} is usually larger than \textsf{SlackQ} over \textsf{Sabre}. However, the Sabre approach performs worse than the Zulehner approach for medium and large benchmarks. It can be seen that \textsf{Zulehner} and \textsf{Sabre} perform well in different scenarios when compared against each other. Regardless, our approach \textsf{SlackQ} outperforms both of them.

\section{conclusion}
The physical layout of contemporary quantum devices imposes limitations for
mapping a high level quantum program to the hardware. It is critical to develop
an efficient qubit mapper in the NISQ era. Existing
studies aim to reduce the gate count but are oblivious to the depth of the
transformed circuit. This paper presents the design of the
first time-efficient slack-aware swap insertion scheme. Experiment results show that our proposed solution generates hardware-compliant circuits with faster execution time
compared with state-of-the-art mapping schemes.

\bibliographystyle{IEEEtran}
\bibliography{ref}

\end{document}